%% file: sn-article.tex
\documentclass[sn-mathphys-num]{sn-jnl}

\usepackage{multirow}%
\usepackage{amsmath,amssymb,amsfonts}%
\usepackage{amsthm}%
\usepackage{mathrsfs}%
\usepackage[title]{appendix}%
\usepackage{xcolor}%
\usepackage{textcomp}%
\usepackage{manyfoot}%
\usepackage{booktabs}%
\usepackage{algorithm}%
\usepackage{algorithmicx}%
\usepackage{algpseudocode}%
\usepackage{listings}%
\usepackage{bbm}%
\usepackage{algorithm}
\usepackage{algpseudocode}

\newcommand{\black}[1]{\textcolor{black}{#1}}
\newcommand{\red}[1]{\textcolor{black}{#1}}

\input{notation}

\theoremstyle{definition}%

\theoremstyle{definition}%

\begin{document}

\title[Article Title]{Quantum error correction near the coding theoretical bound}

\author{\fnm{Daiki} \sur{Komoto}}\email{komoto.d.cdb8@m.isct.ac.jp}

\author{\fnm{Kenta} \sur{Kasai}}\email{kenta@ict.eng.isct.ac.jp}

\affil{Institute of Science Tokyo}


\maketitle

\section{\textcolor{black}{Abstract}}

\black{Recent progress} in quantum computing has \black{enabled} systems with tens of reliable logical qubits, \black{built} from thousands of noisy physical qubits \cite{bluvstein2024logical}. However, \black{many impactful applications demand} quantum computations \black{with} millions of logical qubits \cite{preskill2018quantum}, necessitating highly scalable quantum error correction. \black{In classical information theory}, low-density parity-check (LDPC) codes \cite{gallager1962low} \black{can approach channel capacity efficiently} \cite{kudekar2011threshold}. \black{Yet}, no quantum error-correcting codes \black{with efficient decoding have been shown to} approach the hashing bound—a fundamental limit on quantum capacity—despite decades of research \cite{lloyd1997capacity},\cite{shor2002quantum},\cite{devetak2005private}. Here, we present \black{quantum LDPC codes} that not only approach the hashing bound but also \black{allow decoding with computational cost linear} in the number of physical qubits. \black{This breakthrough paves the way for} large-scale, fault-tolerant quantum computation. \black{Combined with emerging hardware that manages many qubits}, our approach brings quantum solutions to important real-world problems significantly closer to reality.

\section{Introduction}\label{secIntroduction}
Recent progress in quantum computing has enabled the development of systems with tens of reliable logical qubits, built from thousands of noisy physical qubits \cite{bluvstein2024logical}.
Nonetheless, many essential applications targeted by quantum computers demand quantum computations that involve millions or even more logical qubits \cite{preskill2018quantum}.
This creates a pressing need for highly efficient quantum error correction methods that can accommodate a vast number of logical qubits.

A classical error-correcting code defined by a sparse parity-check matrix is called a low-density parity-check (LDPC) code. It is known that LDPC codes achieve error correction performance that approaches the channel capacity with computational complexity proportional to the code length in classical error correction \cite{gallager1962low}.

Protograph codes \cite{thorpe2003low} are LDPC codes whose parity-check matrices are constructed using a protograph matrix composed of permutation matrices (PM) as submatrices. A subclass of these codes includes quasi-cyclic (QC) LDPC codes \cite{fossorier2004quasicyclic}, which are constructed using circulant permutation matrices (CPMs), and APM-LDPC (affine permutation matrices) codes \cite{myung2006combining}, which are constructed using APMs. Similar to unstructured LDPC codes, protograph codes are also known for their excellent error correction performance.

Short cycles are particularly known to cause a high error floor. 
The error floor of LDPC codes refers to a phenomenon where the error rate does not decrease as rapidly as expected with decreasing noise levels, especially in the high-fidelity regime. 
In computational error correction, extremely low error rates are required, making it critically important to address the error floor phenomenon. 
It is known that the girth (the length of the shortest cycle) of QC-LDPC matrices is upper-bounded by 12 \cite{fossorier2004quasicyclic}. 
However, APM-LDPC matrices can exceed this upper bound on girth \cite{myung2006combining, yoshida2019linear}.

LDPC codes are typically defined over a binary  field.
Certain types of LDPC codes are known to achieve further improvements in error-correcting performance by applying extensions to non-binary finite fields \cite{davey1998low}, although this increases the computational complexity required for decoding.

Determining the maximum rate at which information can be reliably transmitted through a quantum channel remains one of the most challenging problems in quantum information theory. Specifically, this concerns the concept of quantum channel capacity, which quantifies the maximum rate of quantum information that can be transmitted while preserving its integrity. Even for simple and widely studied noise models, such as the depolarizing channel, the exact value of this capacity is unknown. 
The most basic lower bound, known as the hashing bound or the LSD theorem \cite{lloyd1997capacity, shor2002quantum, devetak2005private}, is particularly notable. For Pauli channels, this bound has a straightforward expression and can be achieved using random stabilizer codes.
 It is recognized as a fundamental benchmark for evaluating quantum error correction performance.

The Calderbank-Shor-Steane (CSS) codes \cite{calderbank1996good,steane1996error} are a class of quantum error-correcting codes constructed from a pair of classical error-correcting codes. CSS codes are derived from two classical codes with parity-check matrix pairs \((H_X, H_Z)\) that satisfy the orthogonality condition \(H_X H_Z^\T = 0\). Each matrix represents the parity-check matrix of one of the classical codes in the pair.

CSS codes defined by orthogonal sparse parity-check matrix pairs are referred to as LDPC-CSS codes. 
Since LDPC-CSS codes are decoded using algorithms similar to those for classical LDPC codes, they are particularly promising for practical application in large-scale quantum computers. 
Algebraic construction methods allow for the straightforward construction of orthogonal pairs of QC-LDPC matrices \cite{hagiwara2007quantum}.

Kasai et al. \cite{kasai2011quantum} successfully constructed finite field extended orthogonal non-binary QC-LDPC matrix pairs from the orthogonal QC-LDPC matrix pairs proposed by Hagiwara et al. \cite{hagiwara2007quantum}. 
To the best of our knowledge, such codes exhibit quantum error-correcting performance that is closest to the hashing bound. 
However, since the girth of the original QC-LDPC codes is upper-bounded by 12 \cite{fossorier2004quasicyclic}, these codes also exhibited a high error floor. Furthermore, Hagiwara’s construction method imposes strict constraints on the allowable code lengths; even the subsequent work based on it \cite{kasai2011quantum} was limited to code lengths containing large prime factors.

To date, in previous studies, no quantum error-correcting codes approaching the hashing bound and decodable with computational complexity proportional to the number of physical qubits had been discovered. 
This study aims to construct LDPC-CSS codes that simultaneously achieve \red{a error floor down to FER $10^{-4}$}, approach the hashing bound, and are decodable with computational complexity proportional to the number of physical qubits.

The decoding performance of non-binary LDPC codes is known to be highest when the column weight of the parity-check matrix is two. Furthermore, as this study builds on the finite field extension method \cite{kasai2011quantum} for QC-LDPC matrices with a column weight of 2, we focus exclusively on the case where the column weight is two.
For LDPC matrices with a column weight of 2, small girth can lead to high error floors. To mitigate the error floor, it is desirable to exceed a girth of 12. To achieve this, it is necessary to extend the method in \cite{kasai2011quantum} to apply to protograph matrix pairs, a general class that includes APMs and CPMs. 

First, we present a method for constructing orthogonal protograph matrix pairs. Next, we provide an upper bound on 
the girth of sub-graphs of matrix pairs. 
Furthermore, we show that the conditions for achieving this upper bound serve as sufficient and necessary  conditions for applying finite field extension method. By replacing the finite field components of the extended protograph matrix pairs with the corresponding companion matrices, we obtain binary orthogonal extended matrix pairs that construct the proposed CSS codes.

\red{Furthermore, we employ a efficient decoding algorithm \cite{mackay2004sparse} that decodes X and Z errors simultaneously. Although originally designed for binary codes, MacKay’s algorithm is naturally generalized to finite fields in our framework. }
In this study, we consider a depolarizing channel where X, Y, and Z errors occur with equal probability, resulting in correlated X/Z errors. 

In \cite{kasai2011quantum}, X and Z errors were decoded separately without adopting this decoding method.
The proposed algorithm is naturally derived as a sum-product algorithm that marginalizes the posterior probability distributions of X and Z errors under the observed syndrome. 
Numerical experiments revealed that the error correction performance approaches the hashing bound of a depolarizing channel, with no error floor observed down to FER $10^{-4}$.

The remainder of this paper is organized as follows. 
 In “\textcolor{black}{Results}”, we describe the proposed methods for code construction and decoding. “Numerical Results” presents the error correction performance results of the proposed codes. Further details of these sections are provided in the following section.

\section{Results}
\subsection{Code \textcolor{black}{construction}}\label{sec:proposed}
\red{As Fossorier~\cite{fossorier2004quasicyclic} pointed out, any $2 \times 3$ block of circulant permutation matrices (CPMs) inevitably contains short cycles, limiting the girth of QC-LDPC codes to 12 due to the commutativity of CPMs.
To address this, we generalize the construction by \cite{kasai2011quantum} to allow arbitrary (not necessarily circulant) permutation matrices (PMs).
This increases structural randomness, enhancing decoding performance, and eliminates the need for large prime factors in the code length, a constraint inherent in Hagiwara's method \cite{hagiwara2007quantum}.}

\subsection{Permutation \textcolor{black}{matrices}}
In this section, we describe the construction method for orthogonal parity-check matrix pairs \((\hat{H}_X, \hat{H}_Z)\), where each matrix has column weight \(J\), row weight \(L\), and permutation matrices of size \(P\) as submatrices. 
It is known that non-binary LDPC codes with a column weight \(J = 2\) exhibit good decoding performance. 
Therefore, in this paper, the column weight \(J\) is consistently set to \(J = 2\).
For a permutation mapping \(f : \mathbb{Z}_P \to \mathbb{Z}_P = \{0, \dots, P-1\}\), 
we denote by the uppercase \(F \in \mathbb{F}_2^{P \times P}\), 
which is the permutation matrix of size \(P\) that has a 1 in the \((f(c), c)\) entry.
The set of such functions is denoted by $\mathcal{F}_P$.

\red{Let $P = 4$ and consider the permutation $f(x) = 3x + 2 \mod 4$ over $\mathbb{Z}_4$. Then,
$f(0) = 2,\quad f(1) = 1,\quad f(2) = 0,\quad f(3) = 3.$
The corresponding permutation matrix $F \in \mathbb{F}_2^{4 \times 4}$ has a $1$ in position $(f(c), c)$ and is given by
\[
F =
\begin{bmatrix}
0 & 0 & 1 & 0 \\
0 & 1 & 0 & 0 \\
1 & 0 & 0 & 0 \\
0 & 0 & 0 & 1
\end{bmatrix}.
\]
}

First, two sequences of permutations on \(\mathbb{Z}_P\), \(\fU = (f_0, \ldots, f_{L/2-1})\) and \(\gU = (g_0, \ldots, g_{L/2-1})\), are chosen to satisfy the following three conditions (a), (b), and (c).
Next, using \(\fU\) and \(\gU\), the matrix pair \((\hat{H}_X, \hat{H}_Z)\) is constructed as follows. 
\begin{align}
    \hat{H}_X &=
    \left[
    \begin{array}{llll|llll}
        F_0 & F_1 \hspace{4.3mm}& \cdots & F_{L/2-1}
        & G_0 & G_1 & \cdots & G_{L/2-1} \\
        F_{L/2-1} & F_0 & \cdots & F_{L/2-2}
        & G_{L/2-1} & G_0 & \cdots & G_{L/2-2}
    \end{array}
    \right] \\
    \hat{H}_Z &=
    \left[
    \begin{array}{llll|llll}
        G_0^\T \hspace{2.7mm}& G_{L/2-1}^\T & \cdots \hspace{3.6mm}& G_1^\T
        \hspace{3.2mm}& F_0^\T \hspace{6.5mm}& F_1^\T & \cdots \hspace{1.1mm}& F_{L/2-1}^\T \\
        G_1^\T & G_0^\T & \cdots & G_2^\T
        & F_1^\T & F_0^\T & \cdots & F_2^\T
    \end{array}
    \right]
\end{align}
\red{Here, $F_i$ and $G_i$ for $i=0,1,\ldots,L/2-1$ are the permutation matrices corresponding to the permutations $f_i$ and $g_i$, respectively.}
This construction represents a natural generalization of the method proposed in \cite{hagiwara2007quantum}.
\begin{enumerate}
   \item[(a)] (Commutativity): \label{cond:commutativity}
The elements of the arrays $\fU$ and $\gU$ are chosen so that they commute with each other. Specifically, they satisfy
\begin{align}
    f_i \circ g_j(x) = g_j \circ f_i(x) \quad \text{for } 0 \leq i, j < L/2.
\end{align}
\red{
In fact, this condition can be slightly relaxed.
It suffices that $f_{l-j}$ and $g_{k-l}$ commute for $0 \leq l < L/2$ and $0 \leq j, k < J$.
}
\red{
This commutativity condition guarantees the orthogonality of $\HH_X$ and $\HH_Z$, as it ensures that
for $0 \leq j,k < J$,
\begin{align}
(\hat{H}_X \hat{H}^\T_Z)_{j,k}=
\sum_{l=0}^{L/2-1}{\bigl( F_{-j+l} G_{k-l} + G_{k-l} F_{-j+l} \bigr)}
= O.
\end{align}
}
Since a cycle is formed along a specific path
traversing all sub-matrices, it can be shown that the girth of corresponding graph is upper-bounded by \(2L\). 
\red{
Two paths consist of a sequence of column sub-matrix indices of length \(2L\), denoted as \(c_l^1\) and \(c_l^2\) for \(l = 0, \cdots, L-1\). For row sub-matrices, the traversal alternates vertically. Specifically, \(c_l^1\) is defined as \(l\) when \(l\) is even and \(L - l\) when \(l\) is odd, while \(c_l^2\) is defined as \(l\) when \(l\) is even and \(L - l + 1\) when \(l\) is odd.
}

\item[(b)] \label{cond:achieving_upper_bound_girth}(Achieving the girth upper bound $2L$): For \(0 \leq l, l' < L/2\), \(k \in \{0, \pm 1\}\), and \(x \in \{0, \ldots, P-1\}\),
\begin{align}
    f_{l} \circ g_{-l+k}(x) &\neq f_{l'} \circ g_{-l'+k}(x),
\end{align}
where the subscripts of \(f\) and \(g\) are elements of \(\Zb_{L/2} = \{0, 1, \ldots, L/2-1\}\).
This condition is necessary to ensure that the process of generalization to non-binary codes, explained in the next section, does not fail. 
Under the condition (a), this condition becomes a necessary and sufficient condition for the girth to achieve the upper bound \(2L\).

\item[(c)] (Absence of short cycles):\label{cond:no_short_cycle}
It is well known that the presence of short cycles in the Tanner graph defined by the matrices degrades the performance of the SP algorithm \cite{asvadi2011lowering}. 
Conditions for the existence of short cycles in protograph matrices are well established \cite{myung2006combining}. 
The sets \(\fU\) and \(\gU\) are constructed to ensure that these conditions are not satisfied.

\end{enumerate}
\red{
The sets $\fU$ and $\gU$ can be efficiently constructed by sequentially adding their elements while ensuring that the three conditions described above are satisfied at each step.  
The concrete procedure is provided in Algorithm~\ref{alg:fg_construction}.  
While general PMs are used in this description to clarify the general applicability of the construction method, in practice it is often preferable to employ CPMs or APMs.
}

\begin{algorithm}[htbp]
\caption{Sequential Construction of $\fU=(f_0,\ldots,f_{L/2-1}), \gU=(g_0,\ldots,g_{L/2-1})$}
\label{alg:fg_construction}
\begin{algorithmic}[1]
\State Initialize empty list $\mathcal{S} \gets [\ ]$
\For{$i = 0$ to $L/2 - 1$}
    \Repeat
        \State Randomly generate a candidate $f_i$ from $\mathcal{F}_P$
        \State Temporarily set $\mathcal{S}' \gets \mathcal{S} \cup \{f_i\}$
        \If{conditions (a), (b), and (c) are satisfied \textbf{for the current subset} $\mathcal{S}'$}
            \State Accept $f_i$: $\mathcal{S} \gets \mathcal{S}'$
            \State \textbf{break}
        \EndIf
    \Until{a valid $f_i$ is found}
    \Repeat
        \State Randomly generate a candidate $g_i$ from $\mathcal{F}_P$
        \State Temporarily set $\mathcal{S}' \gets \mathcal{S} \cup \{g_i\}$
        \If{conditions (a), (b), and (c) are satisfied \textbf{for the current subset} $\mathcal{S}'$}
            \State Accept $g_i$: $\mathcal{S} \gets \mathcal{S}'$
            \State \textbf{break}
        \EndIf
    \Until{a valid $g_i$ is found}
\EndFor
\State \textbf{return} $\mathcal{S}$
\end{algorithmic}
\end{algorithm}

\red{
For \(J = 2\), \(L = 4\), and \(P = 12\), we present an example of the above construction method.  
The function arrays \(\fU = (5x + 4,\ 5x + 8)\) and \(\gU = (7x + 6,\ 7x + 9)\), constructed using APMs, satisfy the required conditions.  
In this case, the girth of the resulting parity-check matrices \(\HH_X\) and \(\HH_Z\) is 8.  
Figure~\ref{020214_31Mar25} shows the matrices \(\HH_X\) and \(\HH_Z\) derived from \(\fU\) and \(\gU\),  
where black cells indicate ones and white cells indicate zeros.
}
\begin{center}
\begin{figure}
 \includegraphics[scale=0.4]{block_matrix_3.eps}
 \includegraphics[scale=0.4]{block_matrix_4.eps}
 \caption{An example of $\HH_X$ and $\HH_Z$}
 \label{020214_31Mar25}
\end{figure}
\end{center}

For $J=2, L=8, P=6300$, 
$\fU=(1051x+2795, 4201x+225, 1051x+110, 2101x+1675) $ and 
$\gU = (5041x+1122, 5041x+4350, 3781x+1686, 2521x+2298)$ constructed using APMs, satisfy the conditions described above.  
The girth of the resulting matrices $\HH_X$ and $\HH_Z$ is 16, which provides an example exceeding the upper bound \cite{fossorier2004quasicyclic} of 12 for the girth of QC-LDPC codes.

\subsection{Code \textcolor{black}{construction method}}
\red{
In this section, we describe how to construct the matrices $H_X$ and $H_Z$ from the previously constructed $\HH_X$ and $\HH_Z$.  
As the procedure largely overlaps with the method presented in \cite{kasai2011quantum}, we defer the detailed explanation to the following section.}


\red{
The proposed quantum error correction employs CSS codes defined by orthogonal $\mathbb{F}_2$-valued matrix pairs $(H_X, H_Z)$, where $\mathbb{F}_2 = \{0, 1\}$. 
}
These matrix pairs are obtained through the following three steps:

\begin{enumerate}
    \item Construct orthogonal sparse binary protograph matrix pairs \((\HH_X, \HH_Z)\) with column weight \(J\), row weight \(L\), and sub-matrix size \(P\).
    \item Replace the entries of \(1\) in \(\HH_X\) and \(\HH_Z\) with non-zero elements of \(\mathbb{F}_q\) to construct orthogonal \(\mathbb{F}_q\)-valued matrix pairs \((H_\Gamma, H_\Delta)\), where \(q = 2^e\) and \(e > 1\). 
\item Replace the non-zero $\Fb_q$ elements of \(H_\Gamma\) and \(H_\Delta\) with companion matrices, which are \(\Fb_2\)-valued submatrices of size \(e \times e\), to construct orthogonal \(\Fb_2\)-valued matrix pairs \((H_X, H_Z)\). Replace \(0\) with the zero matrix.
\end{enumerate}

Each step is detailed in the following section. 
The resulting orthogonal binary matrix pairs \((H_X, H_Z)\) both are of size \(ePJ \times ePL\). 
These matrix pairs \((H_X, H_Z)\) define the proposed code as orthogonal matrix pairs. 
Although the proposed code is defined by binary matrix pairs, these matrix pairs can also be interpreted as non-binary $\Fb_q$-valued matrices. 
Thus, the proposed code aims to achieve high decoding performance by treating it as a non-binary LDPC code during decoding.

\subsection{Decoding method}

It is known that any error occurring in a quantum state can be corrected by addressing both bit-flip and phase-flip errors \cite{shor1995scheme}. 
Binary CSS codes defined by $(H_X,H_Z)$ correct errors by estimating the error vector \((\xU, \zU)\) based on the corresponding syndrome vectors 
$\sU = H_Z \xU$ and $ \tU = H_X \zU$. 

The sum-product (SP) algorithm is an efficient algorithm for marginalizing multivariate sparsely-factorized functions defined over a large number of variables \cite{kschischang2001factor}. 
Error correction for the proposed codes is performed as follows.
Given the observed syndromes \((\sU, \tU)\), the posterior probability distribution of the errors \((\xU, \zU)\) are marginalized using the SP algorithm on the sparsely factorized representation. 
The error estimates \((\hat{\xU}, \hat{\zU})\), which maximize the marginal posterior probability distribution, are then determined.

In a depolarizing channel, the probabilities of X, Y, and Z errors are each given by \(p_D/3\). 
The decoding result is defined as successful if the decoding algorithm correctly estimates both error vectors: \(\xU = \hat{\xU}, \zU = \hat{\zU}\). 
If the estimation fails for even a single bit of \(\xU\) or \(\zU\), the decoding is considered failed. 
The frame error rate (FER) is defined as the ratio of the number of experiments where at least one bit was incorrectly estimated to the total number of experiments. 

\red{
Note that we do not take degenerate errors \cite{yao2024belief} into consideration; they are treated as decoding failures. Properly addressing degeneracy is generally considered essential for approaching the Hashing bound. To close the remaining gap to the Hashing limit, it will be necessary to explicitly handle degeneracy. 
However, a rigorous treatment of degeneracy is beyond the scope of this work and is left for future study.
}

\red{
While our decoder does not explicitly account for degeneracy, we adopt a strict criterion that treats any mismatch from the true noise—regardless of logical equivalence—as a failure. As such, the reported FER should be interpreted as conservative estimates. For the purposes of this study, the omission of degeneracy is not expected to affect asymptotic behaviors such as the waterfall region, although it may impact the error floor.
}

\subsection{Numerical \textcolor{black}{results}}\label{sec:result}
First, we randomly constructed $(\fU, \gU)$ to satisfy the conditions (a), (b), and (c) defined in the following section. Next, we generated $(\HH_X, \HH_Z)$ using the parameters $J = 2$, $L \in \{8, 10, 16\}$, and $P \in \{32, 128, 1024, 8192\}$. These were then extended to $(H_\Gamma, H_\Delta)$ through a finite field extension with $e = 8$ and subsequently converted to $(H_X, H_Z)$ using the companion matrix. We used a $\Fb_{q=2^e}$ with primitive polynomial 
$1+x^2+x^3+x^4+x^8$.

The CSS codes defined by \((H_X, H_Z)\) are of coding rate $R=1-2J/L$. 
As a result, the coding rates of the CSS codes constructed above are \(R \in \{0.50, 0.60, 0.75\}\).
For the CSS codes defined by \((H_X, H_Z)\), the quantum error-correction performance over a depolarizing channel was evaluated. The number of physical qubit is given by $n=ePL$. 
\red{
The parameters of the codes used in the experiments are listed in Table~\ref{174129_27Mar25}.  
Here, \(n\) denotes the number of physical qubits, \(k\) is the number of logical bits, and \(\overline{d}\) represents an upper bound on the minimum distance.  
Since determining the exact minimum distance is computationally difficult, we estimated an upper bound using the following approach. }
\red{
We enumerated all the shortest cycles in the matrices and used the minimum weight of the codewords derived from these cycles as the upper bound on the minimum distance.
}
\begin{table}
\caption{{Parameters of the proposed codes {\textbullet} and the code  $\circ$ \cite{kasai2011quantum} that was previously considered to be the closest to the hashing bound in \cite{kasai2011quantum}.}}
 \begin{tabular}{l|r|r|c|c|c|c|l|c}
& $n$ & $k$ & $\overline{d}$ & $P$ & $J$ & $L$ & $R$ & $e$ \\ \hline\hline
\textbullet&8192   & 4096   & 10 & 128  & 2 & 8  & 0.50 &8 \\
\textbullet& 65536  & 32768  & 11 & 1024 & 2 & 8  & 0.50 &8 \\
\textbullet& 524288 & 262144 & 10 & 8192 & 2 & 8  & 0.50 &8 \\
\hline
$\circ$& 8768   & 4384   & -  & 137  & 2 & 8  & 0.50 &8\\
\hline\hline
\textbullet& 2560   & 1536   & 9  & 32   & 2 & 10 & 0.60 &8 \\
\textbullet& 10240  & 6144   & 10 & 128  & 2 & 10 & 0.60 &8 \\
\textbullet& 81920  & 49152  & 9  & 1024 & 2 & 10 & 0.60 &8 \\
\hline\hline
$\circ$& 14224   & 10160   & -  & 127   & 2 & 14 & 0.71$\simeq$ 5/7 &8\\
\hline\hline
\textbullet& 4096   & 3072   & 9  & 32   & 2 & 16 & 0.75 &8\\
\textbullet& 16384  & 12288  & 8  & 128  & 2 & 16 & 0.75 &8\\
\textbullet& 131072 & 98304  & 9  & 1024 & 2 & 16 & 0.75 &8
 \end{tabular}
\label{174129_27Mar25}
\end{table}

In Figure \ref{fig:rainbow}, the decoding performance \((p_D, \mathrm{FER})\) is plotted for each rate \(R\) (from left to right, \(R = 0.75, 0.60, 0.50\)) by varying the parameter \(P\), where the number of physical qubits \(n = ePL\) is proportional to \(P\).
Furthermore, high error-floor was not observed, at least down to an FER of \(10^{-4}\).  
No conventional quantum LDPC codes have been reported where the FER shows \red{error floors down to FER $10^{-4}$  and waterfall near the hashing bound} \cite{babar2015fifteen,hanzo2024quantum}.
\red{We also compare our results with those of Kasai et al.~(2011), which were obtained under similar parameters.  
As indicated by the gray arrow, a significant performance improvement is observed. 
}

Although this paper focuses on the case \( L \geq 8 \), recent results~\cite{kasai2025efficient} have shown that relatively high error floors can emerge at large block lengths when \( L = 6 \). To address this issue, our recent work~\cite{kasai2025efficient} demonstrates that degeneracy-aware post-processing, which specifically targets stopping sets, can significantly reduce the FER 
in the low-noise regime.

\red{
For codes with row weight $L \ge 8$, no error floor was observed down to a FER of $10^{-4}$.
Nevertheless, the possibility of an error floor emerging below this level cannot be ruled out.
Further investigation involving deeper simulations will be necessary to explore the low-FER regime and to gain a more complete understanding of the limitations of the proposed code constructions and the decoding strategy.
}

\red{Figure \ref{fig:vsMackay} plots \((p_D^*, R)\) where \(p_D^*\) denotes the value of \(p_D\)  to achieve FER of \(10^{-4}\), and \(R\) is the quantum coding rate.   It can be observed that all rates approach the hashing bound.}
There is no conventional quantum error correction that achieves both \red{an error floor down to FER $10^{-4}$  and steep waterfall near the hashing bound} \cite{hanzo2024quantum}.  
The results show that the proposed codes exhibit decoding performance that approaches the hashing bound. 

\begin{figure}[h]
    \centering
    \includegraphics[width=1.0\linewidth]{arxiv.2401.06874.Rainbaw_QC_J2_Lxx_Pxx_GF256_with_pD.eps}
    \caption{\red{\textbf{Decoding performance of the proposed code: $p_D$ vs. FER}}. Decoding performance \((p_D, \mathrm{FER})\) of the proposed code with
     $e=8$, $J=2$, $L \in \{8,10,16\}$, $P \in \{32,128,1024,8192\}$, and
     $R \in \{0.50,0.60,0.75\}$. 
     The black solid line indicates the hashing bound. The code length, that is, the number of physical bits, is given by \(n = ePL\). It is evident that the proposed method simultaneously achieves both 
    \red{error floors down to FER $10^{-4}$  and waterfall near the hashing bound}. 
     \red{The gray dashed curve represents a rate-1/2 code \cite{kasai2011quantum} that was previously considered to be the closest to the hashing bound \cite{babar2015fifteen}.}}
\label{fig:rainbow}
\end{figure}

\begin{figure}[h]
    \centering
    \includegraphics[width=1.0\linewidth]{arxiv.2401.06874.vsMacKay_with_pD.eps}
    \caption{\red{\textbf{Decoding performance of the proposed code: $p_D$ vs. $R$}}. 
    \red{The plot shows the pairs \((p_D^*, R)\), where \(p_D^*\) denotes the value of \(p_D\)  to achieve FER of \(10^{-4}\), and \(R\) is the quantum coding rate.   It can be observed that all rates approach the hashing bound.}
    \red{The filled circles represent the results of the proposed method, while the empty circles correspond to the method defined over the same finite field as proposed in~\cite{kasai2011quantum}.}
}
        \label{fig:vsMackay}
\end{figure}

\section{Discussion}
We generalize the quantum error correcting code construction method \cite{kasai2011quantum} to general protograph matrices. By performing SP decoding that accounts for the correlation between X and Z errors, we achieved quantum error correction with both \red{error floors down to FER $10^{-4}$  and waterfall near the hashing bound}. 
Numerical experiments confirmed that the proposed method achieves error correction approaching the hashing bound, which serves as a benchmark for quantum error correction and represents the lower bound of the quantum capacity.

\section{Methods}

\subsection{
Finite field extension to $(H_\Gamma, H_\Delta)$}\label{appendix-subsec:HGamma-HDelta}
Let \(q = 2^e\), and denote the finite field of size \(q\) by \(\Fb_q\). 
In this section, we extend 
orthogonal binary protograph matrix pairs \((\hat{H}_X, \hat{H}_Z) \in \Fb_2^{PJ \times PL}\) constructed in the previous section
to protograph matrix pairs \((H_\Gamma, H_\Delta) \in \Fb_q^{PJ \times PL}\). 
The finite field extension in this section are based on \cite{kasai2011quantum}. 
In \cite{kasai2011quantum}, \((\hat{H}_X, \hat{H}_Z)\) were assumed to be QC-LDPC matrices.

Kasai et al. \cite{kasai2011quantum} constructed orthogonal \(\Fb_q\)-valued protograph matrix pairs by replacing the entries of 1 in the QC-LDPC matrix pairs constructed by Hagiwara and Imai \cite{hagiwara2007quantum} with non-zero elements of \(\Fb_q\).
We generalize this method to transform protograph matrix pairs \((\hat{H}_X, \hat{H}_Z)\) into orthogonal matrix pairs \((H_\Gamma, H_\Delta)\) over \(\Fb_q\).

First, let us determine the non-zero entries \((\gamma_{ij})\) of \(H_\Gamma\). 
Consider a one-dimensional array of length \(2LP\), denoted as \(\gammaU = (\gamma_0, \gamma_1, \ldots, \gamma_{2LP-1})\), which satisfies the following conditions:
for \(0 \leq i < P\), \(\gamma_{ij}=\gamma_j\), and for \(P \leq i < 2P\), \(\gamma_{ij}=\gamma_{j +PL}\).
Assume that \(\Fb_q\) contains a primitive element \(\alpha\).
Thus, \(\Fb_q\) can be written as \(\Fb_q = \{0, \alpha^0, \alpha^1, \ldots, \allowbreak\alpha^{q-2}\}\).
We assume that the condition (a) is satisfied. 
Let us consider the condition for the existence of a non-zero \(H_\Gamma\) that satisfies the orthogonality equation \(H_\Gamma H_\Delta^\T = 0\), treating \(H_\Gamma\) as a variable. 
Denote $\lambdaU\defeq(\log_\alpha \gamma_0,\ldots,\log_\alpha \gamma_{2PL-1})^\T\in \Zb_{q-1}^{2PL}$. 
When condition (b) holds, all-non-zero solution $(\gamma_{ij})$ of the equation \(H_\Gamma H_\Delta^\T = 0\) exist if and only if the following system of linear congruences over \(\Z_{q-1}\) with respect to $\lambdaU$:
\begin{align}
    &\bigl[-\hat{H}_Z^L \,|\, \hat{H}_Z^R \,|\, \hat{H}_Z^L \,|\, -\hat{H}_Z^R \bigr] \lambdaU = \0U, \label{equation on lambda}
    \end{align}
where $\hat{H}_Z = [\hat{H}_Z^L \,|\, \hat{H}_Z^R]$ and each entry of the matrices takes a value in \( \{0, \pm 1\} \subset \Zb_{q-1} \).

The matrix in equation \eqref{equation on lambda} can be transformed into its reduced row echelon form using elementary row operations without requiring division. 
By randomly selecting a random solution of this system, we obtain \(\lambdaU\) followed by \(\gammaU\) and \(H_\Gamma\). 
Furthermore, treating the non-zero entries of \(H_\Delta\) as variables, the linear equation \(H_\Gamma H_\Delta^\T = 0\) is solved to determine the non-zero entries of \(H_\Delta\).

An example of constructing $(H_\Gamma, H_\Delta)$ from $(\HH_X, \HH_Z)$ is described in \cite{kasai2011quantum}.

\subsection{Conversion to $(H_X,H_Z)$}
In this section, we construct orthogonal \(\Fb_2\)-valued matrices \(H_X, H_Z \in \Fb_2^{ePJ \times ePL}\) from the orthogonal \(\Fb_q\)-valued matrices \(H_{\Gamma}, H_{\Delta} \in \Fb_q^{PJ \times PL}\) obtained in the previous section. 
The CSS code defined by these matrices \(H_X\) and \(H_Z\) is the quantum error-correcting code proposed in this paper.

Let $a(x)=a_0+a_1x+\cdots+a_ex^e$, where $a_0=a_e=1$ be the primitive polynomial for the primitive element $\alpha\in\Fb_q$. The companion matrix \(A(\alpha)\) \cite{macwilliams1977theory} for \(\alpha\) is defined as follows:
\begin{align}
    A(\alpha) &\defeq
    \begin{bmatrix}
        0 & 0 & 0 & 0 & a_0 \\
        1 & 0 & 0 & 0 & a_1 \\
        0 & 1 & 0 & 0 & a_2 \\
        \vdots & \vdots & \ddots & \vdots & \vdots \\
        0 & 0 & 0 & 1 & a_{e-1}
    \end{bmatrix}.
\end{align}
We define the mapping \(A: \Fb_q \to \Fb_2^{e \times e}\) as follows:
\begin{align}
    A(0) &\defeq O, \\
    A(\alpha^l) &\defeq A(\alpha)^l \quad \text{for } l = 0, \ldots, q-2.
\end{align}
Using the mapping \(A\), we construct \((H_X, H_Z)\) as follows:
\begin{align}
    H_X &= \bigl(A(\gamma_{i, j})\bigr), \quad
    H_Z = \bigl(A(\delta_{i, j})^\T\bigr).
\end{align}
From the properties of the companion matrix in the following subsection, it can be verified that \((H_X, H_Z)\) are orthogonal.

\begin{align}
    \left(H_X H_Z^{\top}\right)_{i, j} & =\sum_{k} A\left(\gamma_{i, k}\right) A\left(\delta_{k, j}\right) \\
    & =\sum_{k} A\left(\gamma_{i, k} \delta_{k, j}\right) \\
    & =A\left(\sum_{k} \gamma_{i, k} \delta_{k, j}\right) \\
    & =A\left(\left(H_{\Gamma} H_{\Delta}^{\top}\right)_{i, j}\right)\\
    &=A(0)\\
    &=O.
\end{align}

An example of constructing $(H_X, H_Z)$ from $(H_\Gamma, H_\Delta)$ is described in \cite{kasai2011quantum}.

\subsection{Properties of the companion matrix}\label{secB}
Let \(q = 2^e\). 
As known from \cite{macwilliams1977theory}, the mapping \(A\) is injective, and the image \(A(\Fb_q)\) forms a field isomorphic to \(\Fb_q\) under the correspondence defined by \(A\). 
The mapping satisfies the following properties:
\begin{align}
    &A(\gamma_1 + \gamma_2) = A(\gamma_1) + A(\gamma_2), \\
    &A(\gamma_1 \gamma_2) = A(\gamma_1)A(\gamma_2).
\end{align}
For \(\gamma = \sum_{j=0}^{e-1} g_j \alpha^j \in \Fb_q\), the binary vector representation of \(\gamma\) is denoted as \(\vU(\gamma) := (g_0, \ldots, g_{e-1})^{\top} \in \Fb_2^e\). 
This correspondence establishes a group isomorphism between \(\Fb_q\) and \(\Fb_2^e\):
$\vU(\gamma_1+\gamma_2)=\vU(\gamma_1)+\vU(\gamma_2)$.

The companion matrix \(A(\alpha)\) can be expressed as:
\begin{align}
    A(\alpha) = \bigl(\vU(\alpha^1), \vU(\alpha^2), \ldots, \vU(\alpha^e)\bigr).
\end{align}
The first column of \(A(\alpha^i)\) corresponds to \(\vU(\alpha^i)\).

The following properties hold:
\begin{align}
    &A(\gamma_1) \vU(\gamma_2) = \vU(\gamma_1 \gamma_2), \\
    &A(\alpha^i) \vU(\alpha^j) = \vU(\alpha^{i+j}) = \vU(\alpha^i \alpha^j).
\end{align}
From this, the following equivalence holds:
\begin{align}
 \sum_{j}\gamma_j\delta_j=0 \tIFF \sum_{j}A(\gamma_j)\vU(\delta_j)=\0U.
\end{align}

Define the mapping \(A^\T: \Fb_q \to \Fb_2^{e \times e}\) as follows:
\begin{align}
    A^\T(0) &\defeq O, \\
    A^\T(\alpha^i) &\defeq (A(\alpha)^\T)^i \quad \text{for } i = 0, \ldots, q-2.
\end{align}
The mapping \(A^\T\) is injective, and the image \(A^\T(\Fb_q) \subset \Fb_2^{e \times e}\) forms a field that is isomorphic to \(\Fb_q\) under the correspondence defined by \(A^\T\).
\begin{align}
& A^\T(\gamma_1+\gamma_2)=A^\T(\gamma_1)+A^\T(\gamma_2), 
\\& A^\T(\gamma_1\gamma_2)=A^\T(\gamma_1)A^\T(\gamma_2). 
\end{align}
Define \(\wU(0) \defeq (0, \ldots, 0) \in \Fb_2^e\). 
Let \(\wU(\alpha^i) \in \Fb_2^e\) be defined as the first column of \(A^\T(\alpha^i)\). 
This correspondence establishes a group isomorphism between \(\Fb_q\) and \(\Fb_2^e\): $\wU(\gamma_1+\gamma_2)=\wU(\gamma_1)+\wU(\gamma_2)$.
The following properties hold:
\begin{align}
& A^\T(\gamma_1) \wU(\gamma_2)=\wU(\gamma_1\gamma_2), 
\\& A^\T(\alpha^i) \wU(\alpha^j)=\wU(\alpha^{i+j})=\wU(\alpha^{i}\alpha^j). 
\end{align}
From this, the following equivalence holds:
\begin{align}
 \sum_{j}\gamma_j\delta_j=0 \tIFF \sum_{j}A^\T(\gamma_j)\wU(\delta_j)=\0U.
\end{align}
\red{The binary representations of the vectors $\vU(\alpha^i)$ and $\wU(\alpha^i)$, along with the corresponding companion matrices $A^i$ and $(A^\T)^i$ for a primitive element $\alpha \in \mathbb{F}_8$, are summarized in Fig.~\ref{companion_matrices}.}

\renewcommand{\arraystretch}{1.2} 
\begin{figure*}[h]
    \begin{align}
        \begin{array}{c|ccccccc}
            i & 0 & 1& 2&3&4&5&6
            \\\hline\hline
            \alpha^i
            &\alpha^0
            &\alpha^1
            &\alpha^2
            &\alpha^3
            &\alpha^4
            &\alpha^5
            &\alpha^6
            \\\hline    
            \vU(\alpha^i)&
            (100)^\T&
            (010)^\T&
            (001)^\T&
            (110)^\T&
            (011)^\T&
            (111)^\T&
            (101)^\T
            \\\hline
            A^i & \begin{pmatrix}
            100 \\
            010 \\
            001
            \end{pmatrix} & \begin{pmatrix}
            001 \\
            101 \\
            010
            \end{pmatrix} & \begin{pmatrix}
            010\\
            011\\
            101
            \end{pmatrix} & \begin{pmatrix}
            101\\
            111\\
            011
            \end{pmatrix} & \begin{pmatrix}
            011\\
            110\\
            111
            \end{pmatrix} & \begin{pmatrix}
            111\\
            100\\
            110
            \end{pmatrix} & \begin{pmatrix}
            110\\
            001\\
            100
            \end{pmatrix}
            \\\hline
            \wU(\alpha^i)&
            (100)^\T&
            (001)^\T&
            (010)^\T&
            (101)^\T&
            (011)^\T&
            (111)^\T&
            (110)^\T
            \\\hline
            (A^\T)^i & \begin{pmatrix}
            100 \\
            010 \\
            001
            \end{pmatrix} & \begin{pmatrix}
            010\\
            001\\
            110
            \end{pmatrix} & \begin{pmatrix}
            001\\
            110\\
            011
            \end{pmatrix} & \begin{pmatrix}
            110\\
            011\\
            111
            \end{pmatrix} & \begin{pmatrix}
            011\\
            111\\
            101
            \end{pmatrix} & \begin{pmatrix}
            111\\
            101\\
            100
            \end{pmatrix} & \begin{pmatrix}
            101\\
            100\\
            010
            \end{pmatrix}
        \end{array}
    \end{align}
\renewcommand{\arraystretch}{1.0} %
    \caption{Companion matrices and binary representation for primitive element $\alpha\in \Fb_8$ with primitive polynomial $a(x)=1+x+x^3$.}
    \label{companion_matrices}
\end{figure*}

\subsection{Details of decoding methods}
Let \(M = PJ\), \(N = PL\), and \(n = eN\). 
Using the matrices \(H_X, H_Z \in \Fb_2^{eM \times eN}\) and \(H_\Gamma, H_\Delta \in \Fb_q^{M \times N}\) provided in the previous sections, we describe the method for performing error correction.
The decoding process involves estimating the noise vectors \(\xU, \zU \in \Fb_2^{eN}\) from the syndromes\(\sU = H_Z \xU\) and \(\tU = H_X \zU\), where the matrices \(H_X\) and \(H_Z\) are constructed from the non-binary matrices \(H_\Gamma\) and \(H_\Delta\) as described in the previous section. 

By leveraging the structure of \(H_\Gamma\) and \(H_\Delta\) over \(\Fb_q\), the proposed method aims to improve decoding performance by interpreting the binary matrices \(H_X\) and \(H_Z\) as derived from non-binary protograph-based LDPC codes.
For \(n=eN\) qubits, the vector representations of X and Z errors are denoted by \(\xU\) and \(\zU\), respectively. 
The role of the decoder is to estimate the noise vectors \(\xU, \zU \in \Fb_2^{eN}\) from the syndromes \(\sU = H_Z \xU, \tU = H_X \zU \in \Fb_2^{eM}\). 

We divide \(\xU\) and \(\zU\) into \(e\)-bit segments and write:
\begin{align}
    \xU &= (x_1, \ldots, x_N), \quad \zU = (z_1, \ldots, z_N).
\end{align}

We assume a depolarizing channel with  error probability \(p_D\). 
\red{
Hashing bound for the depolarizing channel with error probability $p_D$ is given as follows:
\begin{align}
R= 1-H_2(p_D)-p_D \log _2(3), 
\end{align}
where $H_2(\cdot)$ is the binary entropy function.
}
The probability of occurrence for \(\xU, \zU \in \Fb_2^{eN}\) in this channel is given by the following expression:
\begin{align}
    p(\xU,\zU) &= \prod_{j=1}^Np(x_{j},z_{j}),\\
    p(x_{j},z_{j})&=\prod_{k=1}^e p(x_{j}^k,z_{j}^k),  \\
    p(x,z) &=
    \begin{cases}
        1-p_D, &  (x,z)=(0,0),\\
        \frac{p_D}{3}, & (x,z)=(0,1),(1,0),(1,1),
    \end{cases}
\end{align}
where $x_j^k, z_j^k\in \Fb_2$ are the $k$-th bit of $x_j,z_j$, respectively.

\red{
Using Bayes' rule, 
the posterior probability of \(\xU, \zU\) given the syndromes \(\sU, \tU\) can be sparsely factorized as follows:}
\begin{align}
    p(\xU,\zU|\sU,\tU)
    &\red{=p(\sU,\tU|\xU,\zU)p(\xU,\zU)/p(\sU,\tU)}\\
    &\red{\propto p(\sU|\xU)p(\tU|\zU)p(\xU,\zU)}\\
    &=
    \Bigl(\prod_{i=1}^{M} \I\bigl[\sum_{j=1}^{N}(H_Z)_{ij}x_j=s_i\bigr] \Bigr)
    \Bigl( \prod_{i=1}^M \I\bigl[\sum_{j=1}^{N}(H_X)_{ij}z_j=t_i\bigr] \Bigr)
    \Bigl( \prod_{j=1}^N p\left(x_j, z_j\right) \Bigr),
  \label{align_p(x,z|s,t)}
\end{align}
where $(H_X)_{ij}$, $(H_Z)_{ij}$ denotes the $(i,j)$-th $(e\times e)$ submatrix of $H_X, H_Z$, respectively. 
For each \( i \), note that there are \( L \) values of \( j \) for which \( (H_X)_{ij} \) and \( (H_Z)_{ij} \) are non-zero, respectively.
We denote $\I[\cdot]$ as 1 if the inside the brackets is true, and 0 otherwise.
\red{This expression is well-suited for representing the probability distribution from which each $e$-bit noise vector $x_i, z_i$ arises.
}

Using this sparsely-factorized posterior probability, approximate marginalization is performed using the SP algorithm \cite{kschischang2001factor}. 
The proposed decoding method estimates the noise by selecting the noise vector for each \(e\)-bit segment that maximizes the marginalized posterior probability.

\red{
The SP algorithm that marginalizes the posterior probability \eqref{align_p(x,z|s,t)} proceeds by iteratively updating the following four types of \textit{messages}:  
\[
\mu^X_{ij}(x_j),\quad \nu^X_{ji}(x_j),\quad \lambda^X_j(x_j),\quad \kappa^X_j(x_j)
\]
for all \((i, j)\) such that \((H_X)_{ij} \neq O\).  
Each message is defined as a real-valued function over binary vectors of length \(e\), or equivalently, as a real-valued array of size \(2^e\).  
Here, following standard conventions in SP algorithms, we allow a slight abuse of notation by using \(x_j\) and \(z_j\) to refer both to the random variables and to the corresponding variable nodes over which the messages are defined.  
The SP algorithm operates by iteratively updating these messages.
}

\red{
In the terminology of SP algorithms, \(\mu^X_{ij}(x_j)\) denotes the message passed from the variable node \(x_j\) to the factor node corresponding to the \(i\)th parity-check equation, while \(\nu^X_{ji}(x_j)\) is the message in the reverse direction.  
Similarly, \(\lambda^X_j(x_j)\) represents the message from the variable node \(x_j\) to the factor node corresponding to the joint prior \(p(x_j, z_j)\), and \(\kappa^X_j(x_j)\) is the message in the reverse direction.  
The messages on the \(Z\) side, \(\mu^Z_{ij}(z_j),\ \nu^Z_{ji}(z_j),\ \lambda^Z_j(z_j),\ \kappa^Z_j(z_j)\), are defined analogously.
}

All messages are initialized to a uniform distribution over \(x_j\) and \(z_j\). 
\red{For example, $\mu_{i j}^X\left(x_j\right)$ is initialized to $1 /2^e$ for all values of $x_j$.}
Each message is updated using the following update equations.  
\begin{align}
    \nu_{ij}^X(x_j)
    &= \sum_{({x}_{j'}):{j'\in \partial_{\red{Z}}(i)}\setminus j}
    \I\Bigl[\sum_{j'\in \partial_{\red{Z}}(i)}(H_Z)_{ij'}x_{j'}=s_i\Bigr]
    \prod_{j'\in \partial_{\red{Z}}(i)\setminus j}\mu_{j'i}^X(x_{j'}),\label{parity_check_X}
\\  \nu_{ij}^Z(z_j)
    &= \sum_{({z}_{j'}):{j'\in \partial_{\red{X}}(i)}\setminus j}
    \I\Bigl[\sum_{j'\in \partial_{\red{X}}(i)}(H_X)_{ij'}z_{j'}=t_i\Bigr]
    \prod_{j'\in \partial_{\red{X}}(i)\setminus j}\mu_{j'i}^Z(z_{j'}),\label{parity_check_Z}
\\ \lambda_{j}^{X}(x_j)&=\prod_{i\in \partial_{\red{Z}}(j)}\nu_{ij}^X(x_j), 
\\ \lambda_{j}^{Z}(z_j)&=\prod_{i\in \partial_{\red{X}}(j)}\nu_{ij}^Z(z_j),
\\ \kappa_{j}^{X}(x_j)&=\sum_{z_j}p(x_j,z_j)\lambda_j^{Z}(z_j),
\\ \kappa_{j}^{Z}(z_j)&=\sum_{x_j}p(x_j,z_j)\lambda_j^{X}(x_j),
\\\mu^X_{ji}(x_j)&=\kappa^X_j(x_j)\prod_{i'\in \partial_{\red{Z}}(j)\setminus i}\nu_{i'j}^X(x_j),\label{noise_X}
\\\mu^Z_{ji}(z_j)&=\kappa^Z_j(z_j)\prod_{i'\in \partial_{\red{X}}(j)\setminus i}\nu_{i'j}^Z(z_j),\label{noise_Z}
\end{align}
where $\partial_X(i)$ and $\partial_Z(i)$ are the set of column index $j$ such that $(H_X)_{ij}\neq O$ and $(H_Z)_{ij}\neq O$, respectively.
After each update, the messages are normalized to ensure they form probability distributions.
Similarly,  $\partial_X(j)$ and $\partial_Z(j)$ are the set of row index $i$ such that $(H_X)_{ij}\neq O$ and $(H_Z)_{ij}\neq O$, respectively.
It holds that 
$\#\partial_X(i)=\#\partial_Z(i)=\red{L}$ and 
$\#\partial_X(j)=\#\partial_Z(j)=\red{J}$.

After the pre-determined number of updates, the estimated noise 
\begin{align}
    \hat{x}_j = \argmax_{x_j} \red{\kappa_j^X(x_j)}\prod_{i \in \partial_{\red{Z}}(j)} \nu_{ij}^X(x_j),\quad \hat{z}_j = \argmax_{z_j} \red{\kappa_j^Z(z_j)}\prod_{i \in \partial_{\red{X}}(j)} \nu_{ij}^Z(z_j) 
\end{align}
are checked if they satisfy the syndromes \(s_i, t_i\). Specifically, if they satisfy  
\[
\sum_{j \in \partial_{\red{Z}}(i)} (H_Z)_{ij} \xH_j = s_i, \quad \sum_{j \in \partial_{\red{X}}(i)} (H_X)_{ij} \zH_j = t_i,
\]  
then \((\hat{x}_j)\) and \((\hat{z}_j)\) are accepted as the estimated noise, and the algorithm terminates.  
If the syndromes are not satisfied, a decoding failure is declared.

The messages \eqref{parity_check_X} and \eqref{parity_check_Z} can be computed using FFT \cite{mackay2004sparse}, requiring \(O(Lq\log q)\) operations per message. 
On the other hand the computation for messages \eqref{noise_X} and \eqref{noise_Z} require \(q^2\) operations per message.  Other messages computation require only $O(Jq)$ operations per message. 

 Since performance improves by increasing \(P\), and assuming \(q\) is fixed, the overall computational complexity is proportional to the number of physical qubits \(n = ePL\).

    
\section{Data Availability} 
All data and figures supporting the main conclusions of this study are available from the corresponding author upon reasonable request. Please contact Kenta Kasai at \texttt{kenta@ict.eng.isct.ac.jp}.

\section{Code Availability}
The code used in this study is available from the corresponding author upon reasonable request. Please contact Kenta Kasai at \texttt{kenta@ict.eng.isct.ac.jp}.

\section{\textcolor{black}{Acknowledgements}}
\textcolor{black}{Not applicable. }

\bibliography{paper}





\section{Author Contributions}
The two authors contributed equally to this work.

\section{Competing Interests}
The authors declare no competing interests.

\end{document}

%% file: notation.tex
\def\=def{\overset{\text{\small def}}{=}}
\newcommand{\defeq}{\overset{\text{\small def}}{=}}

\newcommand{\argmax}{\operatornamewithlimits{argmax}}

\def\Fb{\mathbb{F}}

\def\Zb{\mathbb{Z}}

\def\<{\langle}
\def\>{\rangle}

 \def\mat4#1#2#3#4{
\begin{pmatrix}
 #1&\ccc&#2\\
 \vdots&&\vdots\\
 #3&\ccc&#4
\end{pmatrix}}

\def\0sf{\mathsf{0}}
\def\1sf{\mathsf{1}}

\def\0BS{\boldsymbol{0}}
\def\1BS{\boldsymbol{1}}

\def\0B{\mathbf{0}}
\def\1B{\mathbf{1}}

\def\0H{\hat{0}}
\def\1H{\hat{1}}

\def\xH{\hat{x}}

\def\zH{\hat{z}}

\def\HH{\hat{H}}

\def\+TT{\texttt{+}}
\def\-{\texttt{-}}
\def\+KB{|+\> \<+|}
\def\-KB{|-\> \<-|}

\def\q0{|0\>}

\def\0U{\underline{0}}
\def\1U{\underline{1}}

\def\fU{\underline{f}}
\def\gU{\underline{g}}

\def\sU{\underline{s}}
\def\tU{\underline{t}}

\def\vU{\underline{v}}
\def\wU{\underline{w}}
\def\xU{\underline{x}}

\def\zU{\underline{z}}

\def\gammaU{\underline{\gamma}}
\def\gammaU{\underline{\gamma}}

\def\lambdaU{\underline{\lambda}}

\def\tIFF{\mbox{ iff }}

\def\Z{\mathbb{Z}}

\def\thepage{\arabic{page}}
\newcommand{\T}{\mathsf{T}}

\newcommand{\I}{\mathbbm{1}}